\documentclass[prb,twocolumn,aps,superscriptaddress,showpacs]{revtex4-1}
\usepackage{graphicx}
\usepackage{color}
\begin{document}
\title{Two distinct topological phases in the mixed valence compound YbB$_6$ and its differences from SmB$_6$}
\author{Tay-Rong Chang}
\affiliation{Department of Physics, 
National Tsing Hua University, 
Hsinchu 30013, Taiwan}

\author{Tanmoy Das}
\affiliation{Centre for Advanced 2D Materials and Graphene Research Centre
National University of Singapore, Singapore 117546}
\affiliation{Department of Physics, National University of Singapore, Singapore 117542}

\author{Peng-Jen Chen}
\affiliation{Department of Physics, National Taiwan University, Taipei 10617, Taiwan}
\affiliation{Nano Science and Technology Program, Taiwan International Graduate Program, Academia Sinica, Taipei 11529, Taiwan and National Taiwan University, Taipei 10617, Taiwan}
\affiliation{Institute of Physics, Academia Sinica, Taipei 11529, Taiwan}

\author{Madhab Neupane}
\affiliation{Joseph Henry Laboratory and Department of Physics,
Princeton University, Princeton, New Jersey 08544, USA}

\author{Su-Yang Xu}
\affiliation{Joseph Henry Laboratory and Department of Physics,
Princeton University, Princeton, New Jersey 08544, USA}

\author{M. Zahid Hasan}
\affiliation{Joseph Henry Laboratory and Department of Physics,
Princeton University, Princeton, New Jersey 08544, USA}

\author{Hsin Lin}
\email{nilnish@gmail.com}
\affiliation{Centre for Advanced 2D Materials and Graphene Research Centre
National University of Singapore, Singapore 117546}
\affiliation{Department of Physics, National University of Singapore, Singapore 117542}

\author{Horng-Tay Jeng}
\affiliation{Department of Physics, 
National Tsing Hua University, 
Hsinchu 30013, Taiwan}
\affiliation{Institute of Physics, 
Academia Sinica, 
Taipei 11529, Taiwan}

\author{A. Bansil}
\affiliation{Department of Physics, 
Northeastern University, Boston, 
Massachusetts 02115, USA}
\date{\today}
\begin{abstract}
We discuss the evolution of topological states and their orbital textures in the mixed valence compounds SmB$_6$ and YbB$_6$ within the framework of the generalized gradient approximation plus onsite Coulomb interaction (GGA+U) scheme for a wide range of values of $U$. In SmB$_6$, the topological Kondo insulator (TKI) gap is found to be insensitive to the value of $U$, but in sharp contrast, Kondo physics in isostructural YbB$_6$ displays a surprising sensitivity to $U$. In particular, as $U$ is increased in YbB$_6$, the correlated TKI state in the weak-coupling regime transforms into a $d-p$-type topological insulator phase with a band inversion between Yb-5$d$ and B-2$p$ orbitals in the intermediate coupling range, without closing the insulating energy gap throughout this process. Our theoretical predictions related to the TKI and non-TKI phases in SmB$_6$ and YbB$_6$ are in substantial accord with recent angle-resolved photoemission spectroscopy (ARPES) experiments.  
\end{abstract}
\maketitle
\section{INTRODUCTION}

Topological insulators (TIs) present a fundamentally new state of quantum matter \cite{TI-1,TI-2,TI-3,TI-4}, which harbors a non-trivial bulk electronic structure with gapless surface states on the boundaries. Experimental discoveries followed soon after the first theoretical proposals in both two-dimensional (2D)\cite{TI-1,3DTI_Kane,Bi2Se3-2,HgTe-1} and 3D materials\cite{Bi2Se3-2} were made. Subsequently, many distinct classes of TIs have been proposed, including: topological Mott insulators\cite{TMI,Mott}; topological Anderson insulators\cite{TAIs}; topological crystalline insulators\cite{TCIs}; and topological Kondo insulators (TKIs) \cite{Kondo-1,Kondo-2}. A variety of broken symmetry TIs have also been proposed including topological antiferromagnetic insulators\cite{TAFIs}, topological hidden order\cite{THO}, and fully gapped topological superconductors.\cite{TI-2,TI-4} Among these many possibilities, TKIs  have attracted much recent attention because of their potential realization in SmB$_6$\cite{SmB6-transport-1,SmB6-transport-2,SmB6-transport-3, SmB6-ARPES-1,SmB6-ARPES-2,SmB6-ARPES-3,SmB6-ARPES-4,SmB6-ARPES-5,SmB6-ARPES-6,SmB6-STM} and YbB$_6$ \cite{YbB6-ARPES-1,YbB6-ARPES-2,YbB6-ARPES-3}, some inconsistencies between the predicted and experimentally observed properties in these systems notwithstanding.

Several lanthanide and actinide materials that contain localized $f$-electron have been predicted to be topologically nontrivial.  
For example, actinide based 5$f$ materials are predicted to be TIs when onsite Coulomb interaction is included in the computations to account for strong electron correlation effects.\cite{Actinide} Plutonium based 5$f$ compounds and the isostructural samarium and ytterbium based 4$f$ compounds are frequently considered to be TKIs\cite{PuB6,Kondo-1,Kondo-2,SmB6-Feng}. 
The conundrum of the TKIs arises from the complex interplay between many energy scales which involve hybridization between localized $f$-orbitals with conduction bands, strong Coulomb interactions, spin-orbit coupling (SOC), and hierarchy of band orderings driven by the crystal environment.
In fact, the present first-principles study reveals that when we consider a wide range of values of $U$, the TKI state survives in SmB$_6$, but in sharp contrast, YbB$_6$ displays a complex evolution of different topological phases with varying interaction strength. In particular, when finite $U$ is introduced in the computations on YbB$_6$, the 4$f$ states are removed from the Fermi energy (E$_F$), consistent with recent ARPES results\cite{YbB6-ARPES-1,YbB6-ARPES-2,YbB6-ARPES-3}; the system transforms from being a TKI into a new $d-p$-type non-trivial TI in the intermediate range of $U$, and finally into a trivial band insulator at large value of $U$.

Theoretical predictions of a TKI state in SmB$_6$\cite{Kondo-1,Kondo-2,SmB6-Feng} were followed by experimental confirmations from transport \cite{SmB6-transport-1,SmB6-transport-2,SmB6-transport-3}, ARPES\cite{SmB6-ARPES-1,SmB6-ARPES-2,SmB6-ARPES-3,SmB6-ARPES-4,SmB6-ARPES-5,SmB6-ARPES-6}, and scanning tunneling microscopy/spectroscopy (STM/STS)\cite{SmB6-STM}. In this way, the existence of an odd number of in-gap surface states has been established in SmB$_6$. Also found is a 2D conductance channel at low temperatures, giving insight into the puzzle of why a residual resistivity persists in SmB$_6$ even at very low temperatures.\cite{SmB6_early_resistivity} However, theoretical interpretation of these metallic surface states still remains controversial\cite{SmB6_nonTKI, SmB6_review}.

SmS\cite{SmS} and the rare-earths YbB$_6$ and YbB$_{12}$\cite{YbB6-Feng} have also been predicted to harbor the TKI state. Early transport measurements indicated that YbB$_6$ is a doped semiconductor\cite{YbB6-transport}. Recent ARPES experiments \cite{YbB6-ARPES-1,YbB6-ARPES-2,YbB6-ARPES-3} however find that the Yb-4$f$ states lie $\sim$ 1 eV below the E$_F$, suggesting a non-TKI phase, contradicting theoretical predictions. On the other hand, ARPES has observed linearly dispersing quasi-2D states without a detectable out-of-the-plane dispersion at the time-reversal invariant momenta $\Gamma$ and X, which is consistent with the presence of an odd number of surface states crossing the Fermi level as expected in a Z$_2$ TI\cite{YbB6-ARPES-1,YbB6-ARPES-2,YbB6-ARPES-3}. The temperature and photon energy dependence of the spectrum\cite{YbB6-ARPES-1}, circular dichroism photoemission spectra \cite{YbB6-ARPES-2}, and spin-resolved ARPES\cite{YbB6-ARPES-3} provide further evidence to support the notion that YbB$_6$ is a non-trivial TI.

In this study, we examine the nature of electronic states in SmB$_6$ and YbB$_6$, which share the same crystal structure but display very different topological phases. To this end, we delineate how the topological phases in these two materials evolve comparatively when the hybridization between the $f$ and conduction levels is turned on or off, as well as when the strength of the on-site Coulomb interaction $U$ is varied. By analyzing the characters of the orbitals involved, we show that the insulating gap in SmB$_6$ indeed stems from the interaction between the localized Sm-4f and itinerant Sm-5d bands. A parity analysis supports the presence of a TKI ground state. Furthermore, the topological nature of SmB$_6$ is found to be essentially independent of the strength of electron-electron interaction. On the other hand, the electronic structure of YbB$_6$ is qualitatively similar to that of SmB$_6$ only in the small $U$ or weak-coupling regime. As $U$ is increased, the TKI phase in YbB$_6$ transforms into a $d-p$-type TI phase. In the intermediate coupling range, the low energy bands around E$_F$ are composed of Yb-5d and B-2p orbitals, and an inverted gap opens up due to the SOC. With further increase in $U$, the preceding band inversion between the 5$d$ and 2$p$ orbitals is removed, and a trivial state is restored in YbB$_6$ but not in SmB$_6$. We are not aware of a previously proposed $d-p$-type TI in the literature.

The remainder of this article is organized as follows. In Sec.~II, we discuss the crystal structure of SmB$_6$ and YbB$_6$ and the methodological details of our first-principles calculations. Section III discusses the electronic structure of SmB$_6$. Sec.~IV turns to consider the electronic structure and the topological phase diagram of YbB$_6$ as a function of the strength of electron-electron interactions. A few concluding remarks are made in Sec.~V.

%%%%%%%%%%%%%%%%%%%%%%%%%%%%%%%%%%%%%%%%%%%%%%%%%%%%%%%%%%%%%%%%%%%%%%%%%%

\section{COMPUTATIONAL DETAILS}
% Crystal structure

%-----------------------------------------------------------------------------------
\begin{figure}[!t]
\includegraphics[width=8.6cm]{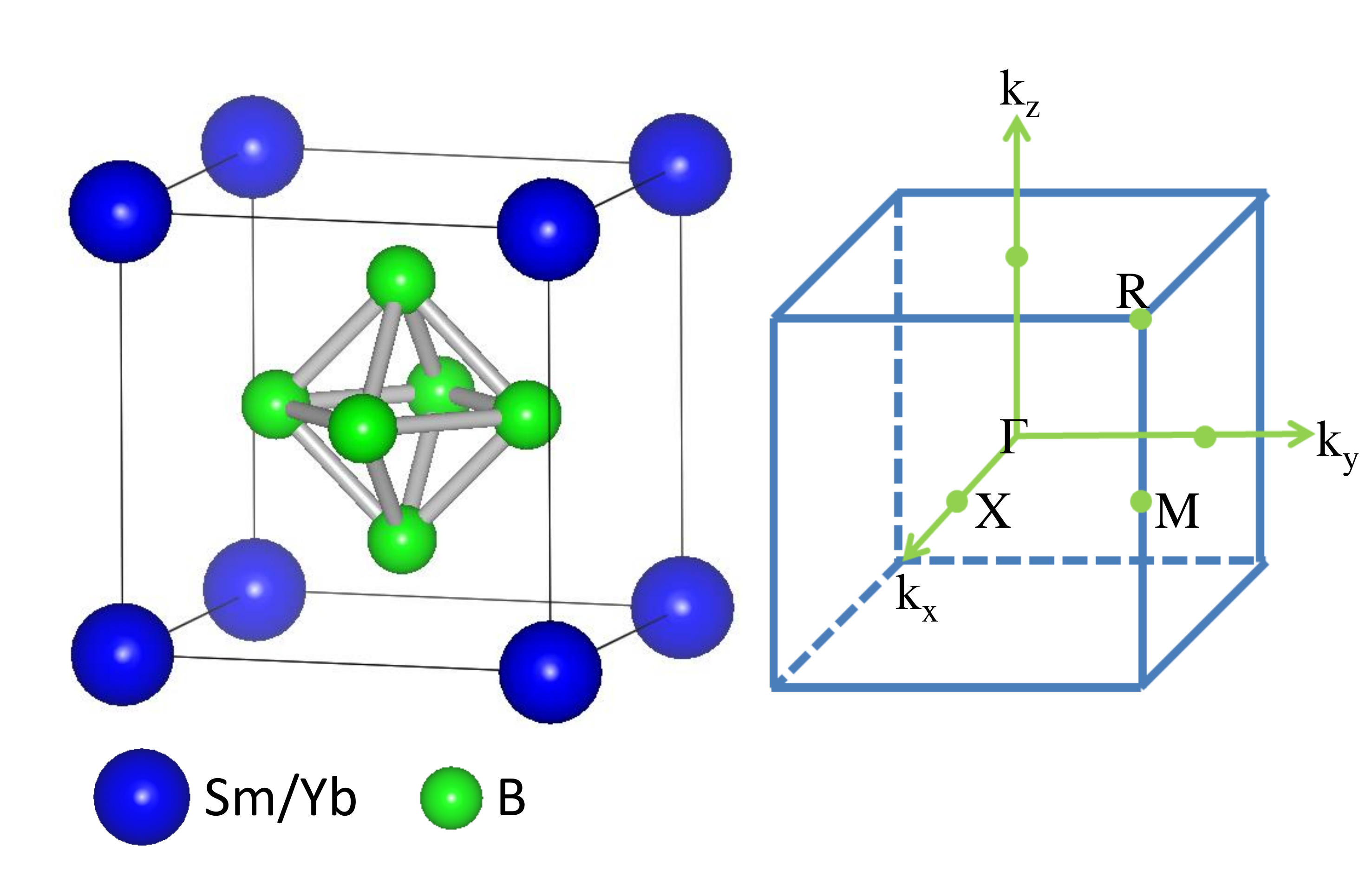}
\caption{ Crystal structure and the first BZ of Sm(Yb)B$_6$.
}
\label{structure}
\end{figure}
%------------------------------------------------------------------------------------

Both SmB$_6$ and YbB$_6$ crystallize in the CsCl-type structure with the Sm or Yb ions and the B$_6$ octahedral clusters lying at the corners and the body centers of the cubic lattice, respectively, see Fig. \ref{structure}. The bulk Brillouin zone (BZ) is cubic, where the center of the BZ is the $\Gamma$-point, diagonal is the R-point, the center of each edge is the M-point, and the center of each face is the X-point. There are three equivalent X(M) points as a result of the cubic symmetry of the crystal. We computed electronic structures using the projector augmented wave method \cite{PAW-1,PAW-2} as implemented in the VASP \cite{VASP-1,VASP-2,VASP-3} package within the generalized gradient approximation (GGA) \cite{PBE} and GGA+U \cite{GGAU} schemes. Experimental lattice constants were used for both SmB$_6$\cite{SmB6-str} and YbB$_6$\cite{YbB6-str}. A 12 $\times$ 12 $\times$ 12 Monkhorst-Pack k-point mesh was used in the computations. The SOC effects are included self-consistently.

%%%%%%%%%%%%%%%%%%%%%%%%%%%%%%%%%%%%%%%%%%%%%%%%%%%%%%%%%%%%%%%%%%%%%%%%%%%%

\section{Electronic structure of ${\rm SmB}_6$}

%----------------------------------------------------------------
\begin{figure}[!t]
 \includegraphics[width=8.6cm]{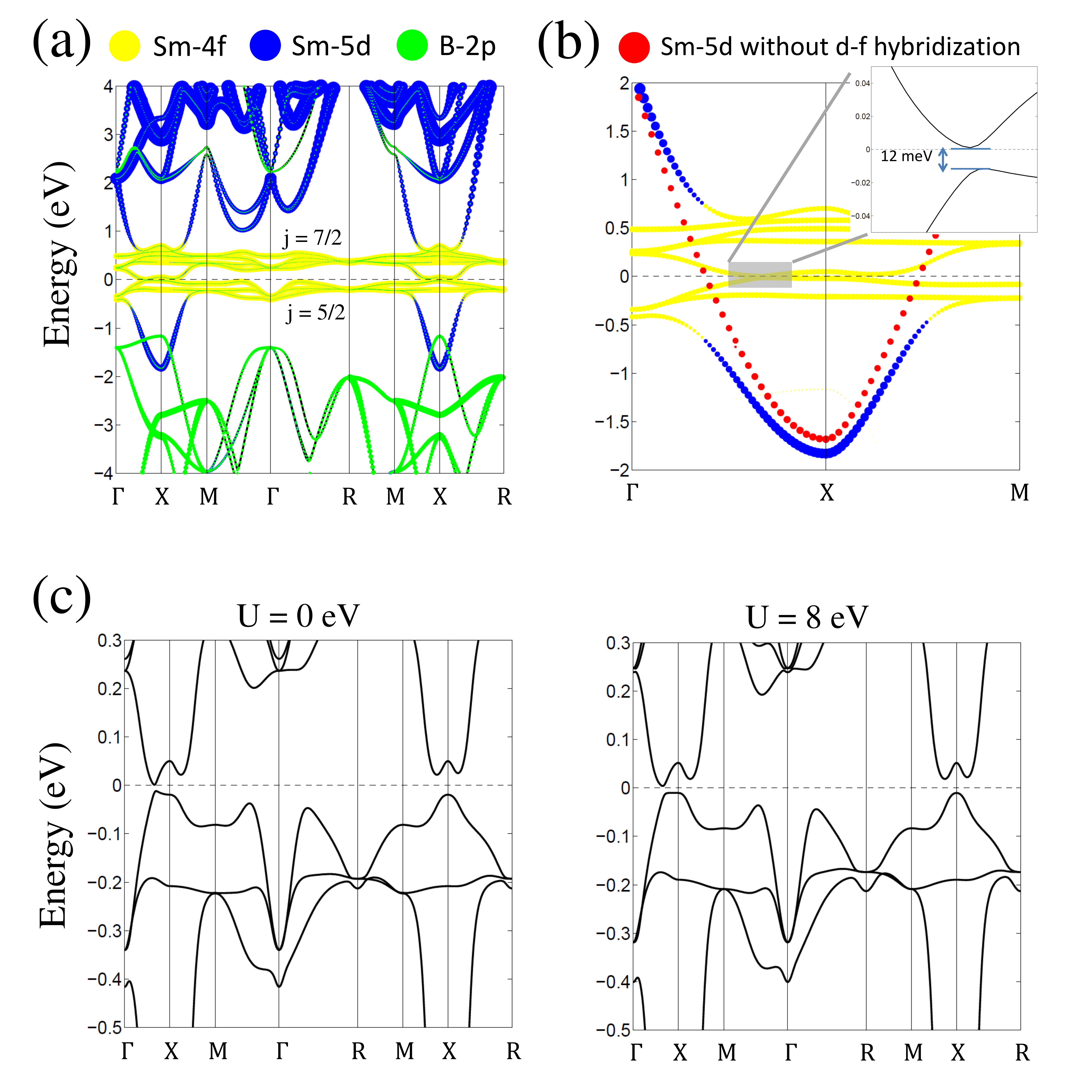}
\caption{ (a) Band structure of SmB$_6$ from GGA calculations.
Sizes of yellow, blue, and green dots denote weights of Sm-4f, Sm-5d, and B-2p orbitals in various bands.
(b) Zoomed in view of (a) near $E_F$. Red dots denote metallic Sm-5d orbitals without hybridization with Sm-4f orbitals.
(c) Low-energy band structure around E$_F$ based on GGA and GGA+U (U = 8 eV) computations.
}
\label{SmB6}
\end{figure}
%----------------------------------------------------------------

Fig.~\ref{SmB6} considers the evolution of electronic structure of SmB$_6$ as a function of $U$. An analysis of the three dominant orbital contributions to the low-lying bands reveals that this system is a Kondo insulator with an inverted band gap of 12 meV lying along the $\Gamma$-X direction, see inset to Fig. \ref{SmB6} (b).\cite{SmB6-Feng,SmB6-Harmon} The Sm-4$f$ orbitals clearly dominate the energy bands over the -0.5 eV to 0.5 eV range, where the bands are split into the $j$ = 5/2 and $j$ = 7/2 states by the SOC (yellow dots in Fig. \ref{SmB6} (a)). In the insulating state, the $j$ = 5/2 states drop below the $E_F$, while the $j$ = 7/2 states remain completely empty. The B-2$p$ states lie around -1.5 eV below E$_F$. According to the ionic configuration of SmB$_6$, Sm is a +2 ion, and thus the Sm-5$d$ states are expected to be empty. In sharp contrast, the Sm-5$d$ states exhibit a surprisingly high degree of itinerancy with a bandwidth as large as 4 eV. The Sm-5$d$ bands span across the $E_F$, and hybridize with the localized Sm-4$f$ states near the $E_F$. The fractional occupations of the Sm-4$f$ and Sm-5$d$ orbitals results in a system with a mixed valence state, which is consistent with various experimental results\cite{SmB6-Mix-1,SmB6-Mix-2,SmB6-Mix-3}.

Fig. \ref{SmB6} shows what happens to the band structure when hybridization between the Sm 4$f$ and 5$d$ states is artificially turned off. The Sm-5$d$ states are now seen to cross $E_F$ without opening an energy gap, thus recovering an ungapped, highly dispersive character (red dots in Fig. \ref{SmB6} (b)). The lower panel of Fig. \ref{SmB6} further examines the stability of Kondo physics over a wide range of $U$ values. Note that the $U$ = 0 limit in a GGA+U computation does not imply total absence of electron correlation effects, and the mixed valence state found for $U$=0 can support Kondo physics, even though the GGA underestimates correlation effects.
Remarkably, little change in topology is seen between $U=0$ and the case where $U$ is as large as 8~eV, an increase in the hybridization gap from 12 meV to about 20 meV notwithstanding. These results demonstrate that SmB$_6$ is a Kondo insulator whose overall band topology is quite immune to electronic interactions as modeled via the on-site Coulomb energy $U$. 

% Parity of SmB6
The band structure of SmB$_6$ shows band inversion around the X-point in Fig. \ref{SmB6} (b) with hybridization turned on. Since SmB$_6$ possesses inversion symmetry, the Z$_2$ topological invariants can be computed via a parity analysis \cite{3DTI_Kane}. The relevant products of parities of occupied wave functions at the time reversal invariant momentum (TRIM) points in the BZ are summarized in Table I. All parity eigenvalues are negative except at the X-points, resulting in Z$_2$ = 1. The system is thus topologically nontrivial with an odd number of gapless surface states residing in the $f$-$d$ hybridization gap. The GGA+U computation for $U$=8 eV is also seen from Table I to yield a topological band structure like the $U$=0 case, albeit with a somewhat larger negative band gap. These results agree with those of Ref. \onlinecite{SmB6-Feng} in predicting that SmB$_6$ harbors a TKI phase.

Recent ARPES experiments observe the presence of an odd number of surface states in SmB$_6$ and give insight into the nature of its insulating gap.\cite{SmB6-ARPES-1,SmB6-ARPES-2,SmB6-ARPES-3,SmB6-ARPES-4}. In particular, Sm-4$f$ orbitals are adduced to lie close to $E_F$, interacting with the highly dispersive Sm-5$d$ orbitals as the temperature is lowered, consistent with theoretical predictions \cite{SmB6-Feng,SmB6-Harmon}. At the same time, the in-gap states are found to develop in the ARPES spectrum at low temperatures \cite{SmB6-ARPES-2,SmB6-ARPES-3,SmB6-ARPES-4,SmB6-ARPES-5,SmB6-ARPES-6}. Fermi surface (FS) mapping reveals the presence of multiple FS pockets: one pocket lies at $\Gamma$ and the other two are centered on the two X-points\cite{SmB6-ARPES-2,SmB6-ARPES-3,SmB6-ARPES-4,SmB6-ARPES-5}. The circular dichroism experiments suggest that the in-gap states possess chirality of the orbital angular momentum \cite{SmB6-ARPES-3}. These observations provide substantial evidence for the existence of topologically protected metallic in-gap states in SmB$_6$. 

%---------------------------------------------------------------------------------
\begin{table}
\label{table1}
\begin{center}
\begin{tabular*}{8cm}{c|cccccc}
\hline \hline
& $\Gamma$ & 3X & 3M & R & Gap (meV) & Type \\
\hline
SmB$_6$ (U = 0 eV) & $\--$ & + & $\--$ & $\--$ & -12 & TKI\\
SmB$_6$ (U = 8 eV) & $\--$ & + & $\--$ & $\--$ & -20 & TKI\\
YbB$_6$ (U = 0 eV) & $\--$ & + & $\--$ & $\--$ & -30 & TKI\\
YbB$_6$ (U = 5 eV) & $\--$ & + & $\--$ & $\--$ & -3 & $d-p$ TI\\
YbB$_6$ (U = 8 eV) & $\--$ & $\--$ & $\--$ & $\--$ & +84 & Trivial \\
\hline \hline
\end{tabular*}
\caption{Products of parities of the occupied states at the eight TRIM points ($\Gamma$, three X, three M, and R) in various cases. The gap at $E_F$ is assigned negative (positive) value for inverted (not inverted) band structure.
} 

\end{center}
\end{table}
%----------------------------------------------------------------------------------
\section{Electronic structure of ${\rm YbB}_6$}

\begin{figure}[!t]
\includegraphics[width=8.6cm]{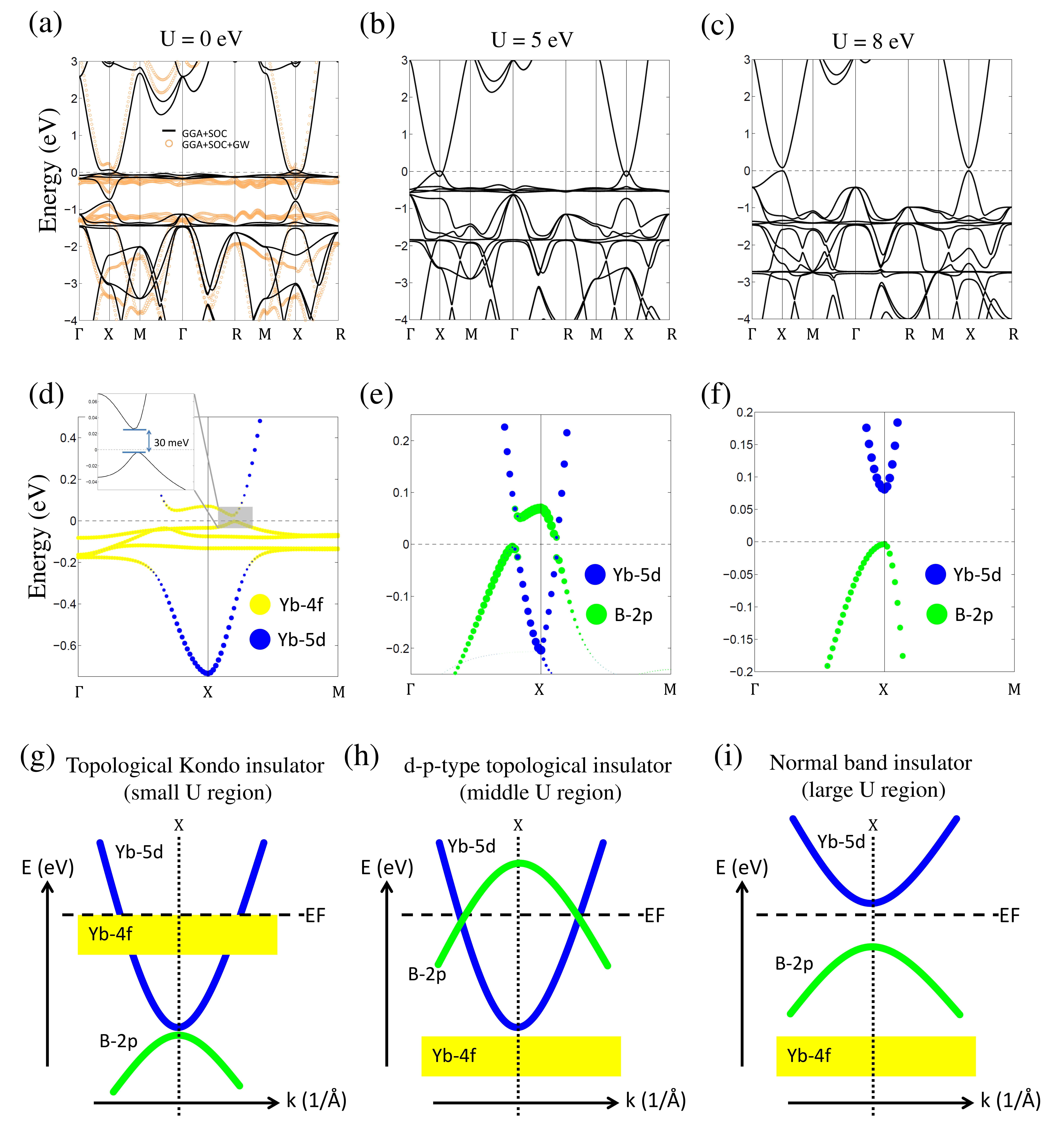}
\caption
{Band structure of YbB$_6$ for (a) $U$ = 0, (b) $U$ = 5 eV, and (c) $U$ = 8 eV. Black solid lines and light orange circles in (a) give  GGA+SOC and GGA+SOC+GW results, respectively. (d-f) Zoom in views of band structures around the $E_F$ corresponding to (a)-(c). Sizes of yellow, blue, and green dots give weights of Yb-4$f$, Yb-5$d$, and B-2$p$ orbitals in the band structures. (g-i) Schematic band structures corresponding to the results of panels (a)-(c).
}
\label{YbB6}
\end{figure}

% Bnad structure of YbB6
Our computed electronic structure for YbB$_6$ is shown in Fig. \ref{YbB6} for three representative values of $U$. At $U$=0, the band structure of YbB$_6$ resembles that of SmB$_6$, although several characteristic differences should be noted. Firstly, unlike SmB$_6$, where the $j$ = 5/2 states lie around E$_F$ and the $j$ = 7/2 states are completely empty (Fig. \ref{SmB6} (a)), the $j$ = 5/2 states of Yb-4$f$ orbitals are fully occupied in YbB$_6$ (-1.5 eV below E$_F$), while the $j$ = 7/2 states lie near $E_F$ (Fig. \ref{YbB6} (a)). Secondly, the insulating band gap is larger in YbB$_6$ with a value of 30 meV along the X-M direction, see inset to Fig. \ref{YbB6} (d). On the other hand, similar to the case of SmB$_6$, the narrow energy gap in YbB$_6$ stems from the hybridization between Yb-4$f$ and Yb-5$d$ orbitals, and the parity analysis of Table I reveals that the system is a TKI for small values of $U$ (Fig. \ref{YbB6} (d)).\cite{YbB6-Feng} Experimental data, however, do not support these theoretical predictions.\cite{YbB6-transport,YbB6-ARPES-1,YbB6-ARPES-2,YbB6-ARPES-3}. Recent ARPES experiments \cite{YbB6-ARPES-1,YbB6-ARPES-2,YbB6-ARPES-3} show that Yb-4$f$ states are located at $\sim$ -1 eV and -2.3 eV below $E_F$, indicating that YbB$_6$ does not belong to the Kondo insulator paradigm. 

%----------------------------------------------------------
%-----------------------------------------------------------

As we turn on electronic interactions via GGA+U calculations, we see substantial changes in the band topology of YbB$_6$ in the two right columns of Fig.~\ref{YbB6}. Yb-4$f$ bands drop below E$_F$ as we increase the value of $U$, with two narrow Yb-4$f$ bands located at -0.5 eV (-1.4 eV) and -1.8 eV (-2.7 eV) binding energy for U = 5 eV (U = 8 eV). As Yb-4f bands move to higher binding energies, their hybridization with conduction bands becomes smaller. The B-2$p$ orbitals, on the other hand, are pushed toward $E_F$ with increasing $U$ and interact with Yb-5$d$ states (Fig. \ref{YbB6} (e)). The low energy band structure clearly shows a band inversion between the Yb-5$d$ and B-2$p$ bands. Combined with the inversion of orbital texture, the parity analysis given in Table I indicates that YbB$_6$ is a non-trivial TI for $U$ values around 5 eV. YbB$_6$ thus presents a novel TI state involving inversion of $d$-$p$ type orbitals.\cite{evenoddTI,SCZhang_review2}  We have recalculated the band structure using a GGA+SOC+GW scheme, and found results similar to those obtained via GGA+SOC calculations, indicating that $d$-$p$ type band inversion cannot be obtained via the GW correction alone, see Fig. \ref{YbB6} (a).

Our preceding scenario is in substantial accord with ARPES data on YbB$_6$, which exhibit nearly linearly dispersing quasi-2D states without a significant out-of-the-plane momentum (k$_z$) dispersion \cite{YbB6-ARPES-1,YbB6-ARPES-2,YbB6-ARPES-3}. Like SmB$_6$, the FS of YbB$_6$ consists of an oval-shaped contour and a nearly circular contour enclosing X- and $\Gamma$-points. The temperature evolution of the states does not reveal a Kondo-like behavior \cite{YbB6-ARPES-1}. Circular dichroism \cite{YbB6-ARPES-2} and spin-resolved ARPES \cite{YbB6-ARPES-3} spectra show that the surface states are spin-polarized and display spin-momentum locking. These experimental results support YbB$_6$ being a TI, but not a TKI.

With further increase of $U$, not only the Yb-4$f$ bands move to higher binding energies, but the band gap between Yb-5$d$ and B-2$p$ states closes and reopens, and the band inversion at the X-points is removed, see Fig. \ref{YbB6} (f). Signs of parity products at all TRIM points are the same for U$>$6.5 eV, and the system thus assumes a trivial band insulator ground state at large U.

%----------------------------------------------------------
\begin{figure}[h]
\includegraphics[width=8.6cm]{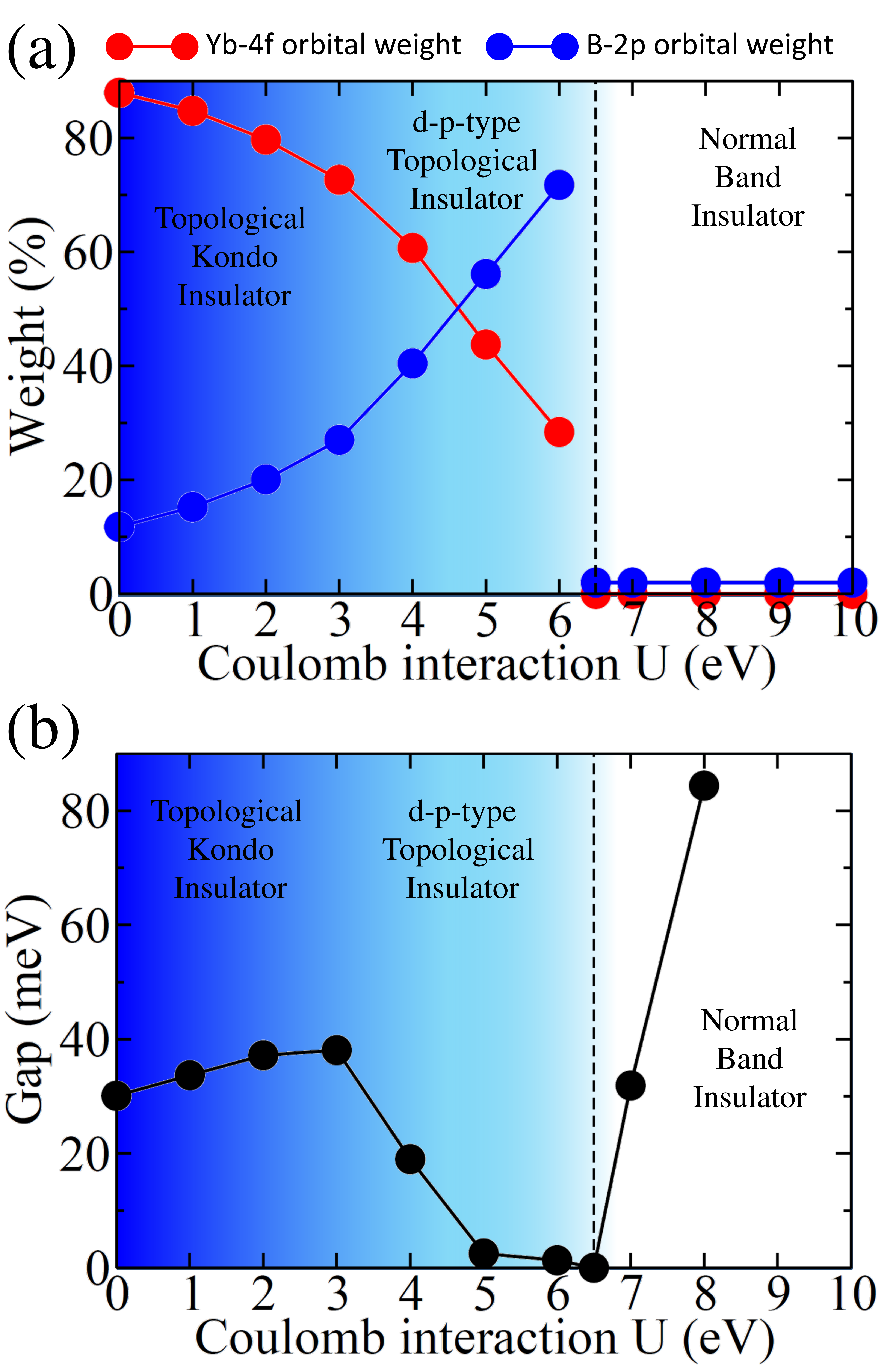}
\caption
{ (a) Weights of Yb-4f (red dots) and B-2p (blue dots) orbitals at the bottom of the conduction band at the X-point as a function of $U$ in YbB$_6$. Lines are drawn through the computed points to guide the eye. (b) Corresponding direct band gaps at the X-point.
}
\label{gwU}
\end{figure}
%-----------------------------------------------------------

Fig.~\ref{gwU} gives insight into the key reasons driving the existence of two distinct nontrivial topological phases in YbB$_6$. Fig.~\ref{gwU}(a) analyzes the contributions of Yb-4$f$ and B-2$p$ orbitals as a function of $U$ at the bottom of the conduction band at the X-point where the band inversion occurs. With increasing $U$, weight of the Yb-4f orbitals decreases, while that of B-2p orbitals increases, indicating that Kondo physics weakens near the $E_F$. The ordering between the Yb-4f and B-2p levels reverses around $U\sim$4 eV, and the  B-2p orbitals start to dominate near $E_F$. This characteristic evolution can also be seen in the band gap in Fig.~\ref{gwU}(b). The band gap at $E_F$ initially increases weakly with increasing $U$ as long as the gap is defined between the Yb-derived $f$ and $d$ bands. Around $U\sim$3~eV, the insulating gap switches over to become a gap between the Yb-5$d$ and  B-2$p$ states, and decreases sharply for higher $U$ values. The system thus undergoes a smooth transition from being a TKI to a $d-p$ type TI without closing the gap. This novel topological phase transition is consistent with adiabatic continuity arguments, which dictate that the topology of two band structures is the same, if one can go from one to the other through a series of adiabatic transformations of the Hamiltonian without closing the gap anywhere in the BZ. YbB$_6$ loses its topological character as the band gap disappears around $U\sim$6.5~eV, and reopens to yield the trivial band insulator at higher values of $U$. 
In this way, modulation of Hubbard U can vary the topological state and lead to a rich phase diagram.\cite{model_tuneU}
Tuning of the Coulomb interaction in a given material is of course not straightforward to realize experimentally. It could perhaps be accomplished, however, via doping with Sm, Pu or other elements with the same crystal structure, which may change the value of effective $U$ through changes in Coulomb screening, and provide a pathway for verifying our predictions.

%\textcolor{blue}{An low-energy tight-binding model calculation with SOC and %dynamical mean-field theory technique also predicted the existence of two %topological phases in such cases, \cite{model_tuneU} however, nature of the %second topological phases in our work is different and novel.} 

\section{CONCLUSIONS}
We have systematically investigated the topological phase diagram of mixed valence compounds SmB$_6$ and YbB$_6$ by varying the strength of on-site Coulomb interaction $U$ within the framework of the GGA+U scheme. The hybridization gap in SmB$_6$ between the localized Sm-4f bands and the conduction bands remains fairly unchanged as $U$ is varied up to values as large as 8~eV. In sharp contrast, YbB$_6$ exhibits a much more complex behavior. As $U$ is increased, Yb-4$f$ states in YbB$_6$ gradually move to lower energies, while the weakly correlated B-2$p$ bands move upward, and these two bands eventually get interchanged without closing the insulating band gap around the $E_F$. In this way, the TKI phase assumed by YbB$_6$ for small (weak coupling) values of $U$ transforms into a new $d$-$p$ type of topological phase for values of $U$ in the intermediate coupling regime. The topological phase in YbB$_6$ reverts to a non-topological state for large values of $U$, but this does not happen in SmB$_6$. 
Our results concerning the topological behavior of pristine SmB$_6$ and YbB$_6$ are in reasonable accord with the available ARPES results on these compounds, and give new insight into the comparative evolution of their band topologies with increasing interaction strength $U$. Our theoretical predictions concerning the novel topological phases of YbB$_6$ might be amenable to experimental testing by substituting Yb by Sm, Pu or other similar elements to drive changes in the value of effective $U$ via changes in Coulomb screening effects in the material. Due to the involvement of $d$ electrons in its topological phase, YbB$_6$ would be prone to quantum fluctuations and instabilities, offering unique opportunities for investigating the robustness of topological protections in the presence of such instabilities.

\section*{ACKNOWLEDGMENTS} 

TRC and HTJ are supported by the National Science Council, Academia Sinica, and National Tsing Hua University, Taiwan. The work at Northeastern University was supported by the US Department of Energy, Office of Science, Basic Energy Sciences contract number DE-FG02-07ER46352, and benefited from Northeastern University's Advanced Scientific Computation Center (ASCC) and the allocation of time at the NERSC supercomputing center through DOE grant number DE-AC02-05CH11231. H.L. acknowledges the Singapore National Research Foundation for support under NRF Award No. NRF-NRFF2013-03. The work at Princeton University is supported by the US Department of Energy grant DE-FG-02-05ER46200. We also thank NCHC, CINC-NTU, and NCTS, Taiwan for technical support.

% Reference


\begin{thebibliography}{99}

% Introduvtion

\bibitem{TI-1}
C. L. Kane and E. J. Mele, 
Phys. Rev. Lett. {\bf 95}, 146802 (2005).

\bibitem{TI-2}
M. Z. Hasan and C. L. Kane, 
Rev. Mod. Phys. {\bf 82}, 3045 (2010).

\bibitem{TI-3}
J. E. Moore, Nature {\bf 464}, 194 (2010).

\bibitem{TI-4}
Xiao-Liang Qi and Shou-Cheng Zhang, 
Rev. Mod. Phys. {\bf 83}, 10571110 (2011).

\bibitem{3DTI_Kane}L. Fu, C. L. Kane, and E. J. Mele, Phys. Rev. Lett. {\bf 98}, 106803 (2007); L. Fu and C. L. Kane, Phys. Rev. B {\bf 76}, 045302 (2007).

\bibitem{HgTe-1}
B.A. Bernevig, T.L. Hughes, and S.C. Zhang, 
Science {\bf 314}, 1757 (2006). 

\bibitem{Bi2Se3-2}
H. Zhang, C.-X. Liu, X.-L. 
Qi, X. Dai, Z. Fang, and S.-C. Zhang, 
Nat. Phys. {\bf 5}, 438 (2009). 

\bibitem{TMI}S. Raghu, X.-L. Qi, C. Honerkamp, and S.-C. Zhang, Phys. Rev. Lett. {\bf 100}, 156401 (2008).

\bibitem{Mott}
D. Pesin and L. Balents, Nat. Phys. {\bf 6}, 376 (2010). 

\bibitem{TAIs} J. Li, R.-L. Chu, J. K. Jain, and S.-Q. Shen, Phys. Rev. Lett. {\bf 102}, 136806 (2009); C. W. Groth, M. Wimmer, A. R. Akhmerov, J. Tworzyd, and C. W. J. Beenakker, Phys. Rev. Lett. {\bf 103}, 196805 (2009).

\bibitem{TCIs}L. Fu, Phys. Rev. Lett. {\bf 106}, 106802 (2011).

\bibitem{Kondo-1}
M. Dzero, K. Sun, V. Galitski, and P. Coleman, 
Phys. Rev. Lett. {\bf 104}, 106408 (2010). 

\bibitem{Kondo-2}
M. Dzero, K. Sun, P. Coleman, and V. Galitski, 
Phys. Rev. B {\bf 85}, 045130 (2012). 

\bibitem{TAFIs}R. S. K. Mong, A. M. Essin, and J. E. Moore, Phys. Rev. B {\bf 81}, 245209 (2010).

\bibitem{THO}T. Das, Sci. Rep. {\bf 2}, 596 (2012).

\bibitem{SmB6-transport-1}
Steven Wolgast, \c{C}a\v{g}lƒ±yan Kurdak, Kai Sun, 
J. W. Allen, Dae-Jeong Kim, and Zachary Fisk,
Phys. Rev. B {\bf 88}, 180405(R) (2013). 

\bibitem{SmB6-transport-2}
D.J. Kim, S. Thomas, T. Grant, J. Botimer, Z. Fisk, Jing Xia,
Scientific Reports {\bf 3}, 3150 (2013).

\bibitem{SmB6-transport-3}
Xiaohang Zhang, N. P. Butch, P. Syers, 
S. Ziemak, Richard L. Greene, and Johnpierre Paglione,
Phys. Rev. X {\bf 3}, 011011 (2013).

\bibitem{SmB6-ARPES-1}
Hidetoshi Miyazaki, Tetsuya Hajiri, Takahiro Ito, 
Satoru Kunii, and Shin-ichi Kimura,
Phys. Rev. B. {\bf 86}, 075105 (2012).

\bibitem{SmB6-ARPES-2}
E. Frantzeskakis, N. de Jong, B. Zwartsenberg, 
Y. K. Huang, Y. Pan, X. Zhang, 
J. X. Zhang, F. X. Zhang, L. H. Bao, 
O. Tegus, A. Varykhalov, A. de Visser, and M. S. Golden,
Phys. Rev. X. {\bf 3}, 041024 (2013).

\bibitem{SmB6-ARPES-3}
J. Jiang, S. Li, T. Zhang, 
Z. Sun, F. Chen, Z.R. Ye, 
M. Xu, Q.Q. Ge, S.Y. Tan, 
X.H. Niu, M. Xia, B.P. Xie, 
Y.F. Li, X.H. Chen, 
H.H. Wen, and D.L. Feng,
Nat. commun. {\bf 4}, 3010 (2013).

\bibitem{SmB6-ARPES-4}
N. Xu, X. Shi, P. K. Biswas, 
C. E. Matt, R. S. Dhaka, Y. Huang, 
N. C. Plumb, M. Radovi$\acute{c}$, J. H. Dil,
E. Pomjakushina, K. Conder, A. Amato, 
Z. Salman, D. McK. Paul, J. Mesot, 
H. Ding, and M. Shi,
Phys. Rev. B. {\bf 88}, 121102(R) (2013).

\bibitem{SmB6-ARPES-5}
M. Neupane, N. Alidoust, S.-Y. Xu, 
T. Kondo, Y. Ishida, D.J. Kim, 
Chang Liu, I. Belopolski, Y.J. Jo,
T.-R. Chang, H.-T. Jeng, T. Durakiewicz, 
L. Balicas, H. Lin, A. Bansil, 
S. Shin, Z. Fisk and M.Z. Hasan,
Nat. commun. {\bf 4}, 2991 (2013).

\bibitem{SmB6-ARPES-6}
Z.-H. Zhu, A. Nicolaou, G. Levy, N. P. Butch, 
P. Syers, X. F. Wang, J. Paglione, 
G. A. Sawatzky, I. S. Elfimov, and A. Damascelli,
Phys. Rev. Lett. {\bf 111} 216402 (2013).

\bibitem{SmB6-STM}
Michael M. Yee, Yang He, Anjan Soumyanarayanan, 
Dae-Jeong Kim, Zachary Fisk, Jennifer E. Hoffman,
arXiv:1308.1085

\bibitem{YbB6-ARPES-1}
Madhab Neupane, Su-Yang Xu, Nasser Alidoust, 
Guang Bian, Dae-Jeong Kim, Chang Liu, 
Ilya Belopolski, Tay-Rong Chang, Horng-Tay Jeng, 
Tomasz Durakiewicz, Hsin Lin, Arun Bansil, 
Zachary Fisk, and M. Zahid Hasan
arXiv:1404.6814

\bibitem{YbB6-ARPES-2}
M. Xia, J. Jiang, Z. R. Ye, 
Y. H. Wang, Y. Zhang, S. D. Chen, 
X. H. Niu, D. F. Xu, F. Chen, 
X. H. Chen, B. P. Xie, T. Zhang, and D. L. Feng,
arXiv:1404.6217

\bibitem{YbB6-ARPES-3}
N. Xu, C. E. Matt, E. Pomjakushina, 
J. H. Dil, G. Landolt, J.-Z. Ma, 
X. Shi, R. S. Dhaka, N. C. Plumb, 
M. Radovic, V. N. Strocov, T. K. Kim, 
M. Hoesch, K. Conder, J. Mesot, 
H. Ding, and M. Shi,
arXiv:1405.0165

\bibitem{Actinide}X. Zhang, H. Zhang, J. Wang, C. Felser, and S.-C. Zhang, Science {\bf 335}, 1464 (2012); 
\bibitem{PuB6}X. Deng, K. Haule, and G. Kotliar, Phys. Rev. Lett. {\bf 111}, 176404 (2013).

\bibitem{SmB6-Feng}
Feng Lu, JianZhou Zhao, Hongming Weng, Zhong Fang, and Xi Dai,
Phys. Rev. Lett. {\bf 110} 096401 (2013).

%\bibitem{TI-HH}H. Lin, L. A. Wray, Y. Xia, S. Xu, S. Jia, R. J. Cava, A. Bansil, and M. Z. Hasan, Nat. Mater. {\bf 9}, 546 (2010); S. Chadov, X. Qi, J. K\82\E0\F6?bler, G. H. Fecher, C. Felser, and S. C. Zhang, Nat. Mater. {\bf 9}, 541 (2010).

\bibitem{SmB6_early_resistivity}J. W. Allen, B. Batlogg, and P. Wachter, Phys. Rev. B {\bf 20}, 4807 (1979); J. C. Cooley, M. C. Aronson, Z. Fisk, and P. C. Canfield, Phys. Rev. Lett. {\bf 74}, 1629 (1995).

\bibitem{SmB6_nonTKI}Z.-H. Zhu, A. Nicolaou, G. Levy, N. P. Butch, P. Syers, X. F. Wang, J. Paglione, G. A. Sawatzky, I. S. Elfimov, A. Damascelli,  Phys. Rev. Lett. {\bf 111}, 216402 (2013).

\bibitem{SmB6_review}J. D. Denlinger, J. W. Allen, J.-S. Kang, K. Sun, B.-Il Min, D.-J. Kim, and Z. Fisk,  arXiv:1312.6636.

\bibitem{SmS}Z. Li, J. Li, P. Blaha, and N. Kioussis, Phys. Rev. B {\bf 89}, 121117 (2014); J.-Z. Zhao, F. Lu, H. Weng, Z. Fang, and X.
Dai, in Bull. Am. Phys. Soc. (American Physical Society, 2014).

\bibitem{YbB6-Feng}
Hongming Weng, Jianzhou Zhao, Zhijun Wang, 
Zhong Fang, and Xi Dai,
Phys. Rev. Lett. {\bf 112} 016403 (2014).

\bibitem{YbB6-transport}
J. M. Tarascon, J. Etourneau, P. Dordor, 
P. Hagenmuller, M. Kasaya and J. M. D. Coey,
J. Appl. Phys. {\bf 51}, 574 (1980).

\bibitem{PAW-1}
 P. E. Bl$\ddot{o}$chl, 
 Phys. Rev. B. {\bf 50}, 17953 (1994).

\bibitem{PAW-2}
 G. Kresse and J. Joubert, 
 Phys. Rev. B. {\bf 59} 1758 (1999).

\bibitem{VASP-1} 
G. Kress and J. Hafner, 
Phys. Rev. B. {\bf 48}, 13115 (1993).

\bibitem{VASP-2} 
G. Kress and J. Furthm$\ddot{u}$ller, 
Comput. Mater. Sci. {\bf 6}, 15 (1996)

\bibitem{VASP-3} 
Phys. Rev. B. {\bf 54} 11169 (1996).

\bibitem{PBE} 
J. P. Perdew, K. Burke, and M. Ernzerhof, 
Phys. Rev. Lett. {\bf 77} 3865 (1996).

\bibitem{GGAU} 
A. I. Liechtenstein, V. I. Anisimov, and J. Zaanen, 
Phys. Rev. B {\bf 52}, R5467 (1995).

\bibitem{SmB6-str}
A. L. Malyshev, V. A. Trounov, D. Y. Chernyshov, 
M. M. Korsukova, V. N. Gurin, O. K. Antson, and
P. E. Hiism$\ddot{a}$ki.
Mater. Sci. Forum {\bf 62}, 69-70 (1990).

\bibitem{YbB6-str}
P. P. Blum and  E. F. Beraut. 
Acta Cryst. {\bf 7}, 81-84 (1954).

\bibitem{SmB6-Harmon}
V. N. Antonov and B. N. Harmon
Phys. Rev. B. {\bf 66}, 165209 (2002).

\bibitem{SmB6-Mix-1}
J.‚ÄâW. Allen, L.‚ÄâI. Johansson, I. Lindau, and S.‚ÄâB. Hagstrom, 
Phys. Rev. B {\bf 21}, 1335 (1980). 

\bibitem{SmB6-Mix-2}
J.‚ÄâN. Chazalviel, M. Campagna, G.‚ÄâK. Wertheim, and P.‚ÄâH. Schmidt, 
Phys. Rev. B {\bf 14}, 4586 (1976). 

\bibitem{SmB6-Mix-3}
E. Beaurepaire, J.‚ÄâP. Kappler, and G. Krill, 
Phys. Rev. B {\bf 41}, 6768 (1990).

\bibitem{evenoddTI}T. Das, Phys. Rev. B {\bf 88}, 035444 (2013).

\bibitem{SCZhang_review2}H. Zhang and S.-C. Zhang, Phys. Status Solidi RRL {\bf 7},72 (2013).

\bibitem{model_tuneU} Jan Werner and Fakher F. Assaad, Phys. Rev. B {\bf 88}, 035113 (2013).


\end{thebibliography}
\end{document}